\RequirePackage{fix-cm}
\documentclass[twocolumn]{svjour3}          
\smartqed  
\usepackage{graphicx}
\usepackage{amsmath}
\usepackage{amssymb}
\usepackage{tikz}
\usepackage{verbatim}
\usepackage{subfig}
\usepackage{array}
\journalname{Software and Systems Modeling}
\begin{document}

\title{Formalizing the Four-layer Metamodeling Stack -- Potential and Benefits
}

\author{Victoria Döller
}

\institute{
				Victoria Döller \at 
				Research Group Knowledge Engineering,\\ 
				Faculty of Computer Science,\\
				University of Vienna, Vienna, Austria\\
				\email{victoria.doeller@univie.ac.at}
}

\date{Received: date / Accepted: date}

\maketitle

\begin{abstract}
	 Enterprise modeling deals with the increasing complexity of processes and systems by operationalizing model content and by linking complementary models and languages, thus amplifying the model-value beyond mere comprehensible pictures.
	 To enable this amplification and turn models into computer-processable structures a comprehensive formalization is needed.
	 This paper presents a generic formalism based on typed first-order logic and provides a perspective on the potential and benefits arising for a variety of research issues in conceptual modeling.
	 We define modeling languages as formal languages with a signature $\Sigma$ -- comprising object types, relation types, and attributes through types and function symbols -- and a set of constraints. 
	 Three cases studies are included to show the effectiveness of the approach. 
	 Applying the formalism to the next level in the hierarchy of models we create \texttt{M2FOL}, a formal modeling language for metamodels.
	 We show that \texttt{M2FOL} is self-describing and therefore complete the formalization of the full four-layer metamodeling stack.	 
	 On the basis of our generic formalism applicable to arbitrary modeling languages we examine three current research topics -- language interleaving \& consistency, operations on models, and automatic translation of formalizations to platform-specific code -- and how to approach them with the proposed formalism.
	 This shows that the rich knowledge stack on formal languages in logic offers new tools for old problems.
\end{abstract}

\section{Introduction}
\label{intro}
Enterprise modeling has proven instrumental in facing the challenges of increasing complexity and interdependences of processes and systems in the modern world.
Research on enterprise modeling has enhanced modeling languages from mere instruments for pictures supporting human understanding to highly specialized tools with value adding mechanisms like information querying, simulation, and transformation \cite{bork19omilab,frank2014biseagenda}. 
The nature of models has evolved from a visual representation of information to an exploitable knowledge structure \cite{buchmann2019educationdesignproblem}. 
Nevertheless the European enterprise modeling community experiences that the potential of enterprise modeling is currently not fully utilized in practice and modeling is employed only by a limited group of experts. Therefore in \cite{sandkuhl16modformasses,sandkuhl2018researchagenda} a research agenda is formulated to establish ``modeling for the masses" (MftM) and broadcast its benefits also to non-experts.

\subsection{A Need for Formalization}
Although the initiators of the MftM movement mention that the formality of model representation possibly hampers understandability, we argue that the idea behind MftM nevertheless requires an investigation of the formal foundations of models and languages. This is for three reasons: 
1) According to the \textit{stakeholder dimension} of challenges in enterprise modeling research \cite[p.234]{sandkuhl16modformasses} computers also have to be seen as stakeholders producing and consuming models. To make models computer-processable they have to be formalized, because as computers do not understand semi-formal or unstructured models and language specifications \cite{bork14formalaspects}. 
2) The vision of models being not autotelic but \textit{being a means to the operationalization of information} \cite[p.229]{sandkuhl16modformasses} calls for value-adding functionality beyond mere graphics like reasoning, verification \& validation or simulation, which is formulated ideally computer-understandably and impl\-ementation-independently, i.e. formalized. 
3) The vision of \textit{local modeling practices which are globally integrative} \cite[p.229]{sandkuhl16modformasses} calls for a common foundation of what models and modeling languages are to enable the linking and merging of models in different domains with different semantics \cite{herrmann2007algebraic}.

Formalization is also essential in the light of the emergent importance of domain specific modeling languages (DSMLs) \cite{frank2013dsml} as well as an increasing agility in the advancement and extension of established languages and methods \cite{karagiannis18amme}. 
The lack of a common way for formalizing DSMLs leads to divergent formal foundations limiting the opportunities to compare or link models.
Frequently the big standards are extended for a specific domain, e.g. the extension of i* with security concepts constituting the modeling language Secure Tropos \cite{mouratidis07securetropos,rrenja2015sectrop}.
Therefore a common way of specifying the base languages as well as the extensions or modules is required.
A silo like formalization of the big standards is not sufficient as divergent base concepts of models and different underlying formal structures can impede a mutual interconnection and integration.

Another important building block for advancing the science of conceptual modeling is an exact and commonly applied method for specifying modeling \linebreak languages. 
A survey conducted by Bork et al. showed that the specification documents of the standardized languages like UML and ArchiMate diverge in the concepts they consider as well as in the techniques they use to specify their visual metamodels \cite{bork20specificationtechniques}.
Examples from recent scientific publications indicate that also in research on domain specific languages no common practice of metamodel specification is in use.
Several contributions specify metamodels with UML class diagrams, declaring object types as classes and relation types as classes or named association arrows, e.g. \cite{zdravkovic21,wimmer20execdsl,ralyte2017evolutionmodels,schoen2019capabilitymod,stirna2017capabdrivdev}. Others simply define the object and relation types with box-and-line models devoid of an underlying language and rely on the intuitive understanding of the reader, e.g. \cite{lara16,paczona19}. 
This shows that although metamodels are models themselves and therefore subject of interest for enterprise modeling research no language for metamodels has been established yet. 
Nevertheless when a language has to be implemented or executed a precise and unambiguous definition of the metamodel is crucial \cite{bork14formalaspects}.

\subsection{Goal and Requirements} \label{secrequirements}
\begin{figure*}
	\centering
	\includegraphics[width=0.8\textwidth]{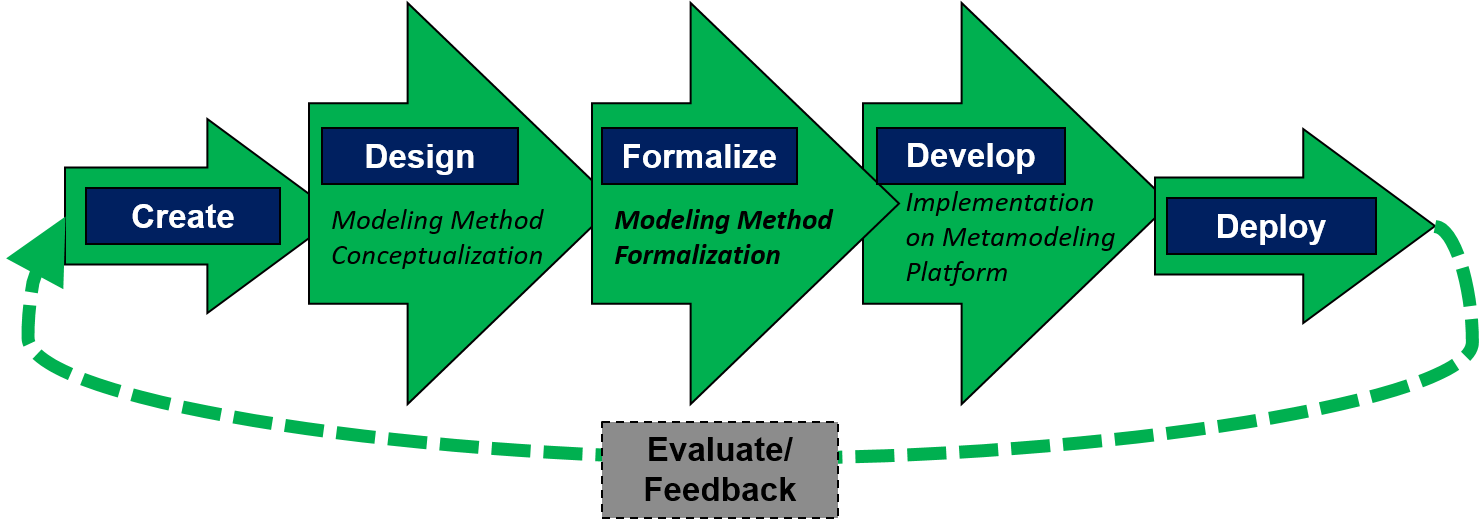}
	\caption{The AMME lifecycle for agile modeling method engineering adapted from \cite{karagiannis18amme}} \label{amme}
\end{figure*}
According to the AMME framework of agile modeling method engineering shown in Figure \ref{amme} the phase of formalization is pivotal in the lifecycle of a modeling language. Yet there is no common procedure of how to formalize arbitrary modeling languages. Often existing formalization approaches restrict to a concrete application, domain or language, thereby limiting the reusability in other domains and languages. As the AMME lifecycle is meant as a generic procedure model for generating arbitrary modeling languages and methods, we need a formalism applicable to any domain and language an engineer might come up with. The work at hand intends to close this gap and aims at building a bridge between the \textit{Design} phase of collecting the relevant concepts and the \textit{Develop} phase of transferring the final design to a metamodeling platform in the AMME lifecycle. 
Such a formalism must be consistent with the semantics and structure of a modeling language. For this reason we have to consider the formal foundations of modeling languages. We summarize the concrete requirements for the formalism as follows: 1) it has to be complete regarding the general building blocks of a language, 2) it has to be faithful to the character of modeling languages as such, and 3) it must be generic in a way that it admits the formalization of any language.

In the work at hand we will demonstrate that in accordance with these requirements modeling languages can be defined as  
as formal languages in the sense of logic. This means they comprise a signature $\Sigma$ for the syntax and a set of constraints, for which we use first-order predicate logic. 
This paper extends our prior work \cite{doeller20} published at the PoEM 2020 conference about the definition of modeling language, where we concretely stated how the core concepts of modeling languages can be expressed in logical terms. 
Predicate logic provides the construct of a $\Sigma$-structure, i.e. an interpretation of the signature, which is the canonical correspondent to the model being an instantiation of a metamodel. 
Applying the definition to the meta language level results in \texttt{M2FOL}, a formal modeling language for metamodels. With \texttt{M2FOL} we are capable of modeling the syntax of a language to be specified, to be more precise the signature of the language according to the definition.
The paper at hand extends the presented definition of formal modeling languages as well as the language \texttt{M2FOL} with the concept of multi-value attributes. We furthermore exemplify the potential and benefits inherent to the proposed formalism on a diverse range of research topics and demonstrate the opportunities established methods from logic can provide for conceptual modeling.

The rest of this paper is structured as follows: In Section \ref{secbackground} we give an overview of related work on formalization of metamodels and modeling languages. In Section \ref{secformalmodelinglanguages} we introduce the definition of formal modeling languages and models and concretize how the basic concepts of a language -- object and relation types, attributes, inheritance, and constraints -- can be expressed in logical terms. We then use this definition in Section \ref{secm2fol} to create \texttt{M2FOL} -- a formal modeling language for metamodels -- and outline its self-de\-scribing character. Given a metamodel specified with \texttt{M2FOL} we show how to algorithmically deduce the signature of the corresponding modeling language. After that we give in Section \ref{secoutlookmethods} an outlook to the potential and benefits of formal modeling languages. We present some ongoing research and approaches on how to interleave modeling languages, formally include operations on models into the language specification, and use the formalization as a single point of specification processable by machines. In Section \ref{seceval} we evaluate the formalism with respect to the formulated goal and requirements and explain the agenda for the empirical evaluation that is currently being conducted.

\section{Background and Related Work}  \label{secbackground}
\subsection{Formalism vs. Formal Syntax}
We begin by discussing the distinction between a \textit{formalism} and a \textit{formal way of specifying a language}. A formalism always gives rise to a formal specification. The converse, however, is not true. 
This can be compared to the concept of a graph and the unique and precise way of specifying a graph with an adjacent matrix. Each graph can be represented as a matrix but a matrix per se does not provide the semantics of graph theory. The same applies to specification techniques merely offering a formal syntax to describe modeling languages. These techniques provide a unique way of specification but lack the structure and semantics of the underlying components of conceptual modeling. 
A formal syntax may offer a notation for specifying inheritance of object types. Nevertheless, the notation does not accomplish the inherent semantics, i.e. the transfer of features of the supertype to the subtype. This behaviour must be added to the syntax system by hand although a suitable underlying structure would entail concepts for inheritance. For this simple case basic set theory would suffice to capture inheritance via sets and containment automatically extending  functions defined on the superset to all subsets.

In current research there is a consensus about a proper underlying structure of modeling languages, that is formal languages as defined in mathematical logic \cite{delcambre18referenceframework,guarino19,olive07conceptual,Partridge13,thalheim11chapter}. Of course not every formal language is a modeling language, but modeling languages form a subclass of all formal languages.
In this paper we want to concretize and work out this class of formal modeling languages. 

Our principal distinction between a formalism and a formal syntax distinguishes our goal from existing approaches like the Meta-Object Facility (MOF) standard which offers a concrete syntax or notation system for specifying modeling languages but no inherent theory and methods for the concepts to be described. Furthermore MOF was developed with metamodels meant as software structure specifications \cite{sprinkle10metamodelling} which restricts metamodeling only to one domain. In contrast to that, our approach regards metamodeling as a generic instrument applicable to a broad range of domains rather than only software engineering.
Also the language Z or the lambda calculus differ from our endeavour in a fundamental way as they are methods developed to support the specification of computation, programming and software development.
Of course they also make use of concepts from logic and may overlap with our approach in some areas. Nevertheless, they do not inherit the semantics of modeling languages and provide no explicit way of describing modeling languages.

\subsection{Formalisms for Concrete Modeling Languages}
According to the Characterizing Conceptual Model Research (CCMR) framework we are interested in contributions located in the dimension \textit{Formalize} working on the level of \textit{Conceptual Modeling Languages} and \textit{Metamodeling Languages} \cite{delcambre19}. 
In this respect, we want to delineate our approach from the various attempts addressing the formalization of a specific modeling language. These attempts mostly aim at supporting a specific purpose or functionality and do not provide means to define arbitrary metamodels and languages. An example is the OSM-logic using typed predicate logic for object-oriented systems modeling with a focus on time-dependent system behaviour \cite{clyde12osmlogic}. 
Another example is the SAVE method for simulating IoT systems using the $\delta$-calculus \cite{lee2016save}. 
These specific formalizations may offer ideas suitable for being generalized to a generic approach but will not be comprehensively discussed here. 
However, as soon as there is a common practice of formally defining the ubiquitous concepts of modeling languages these specific approaches can be constructed as reusable extensions and modules and be of value in a broader field of application.

\subsection{Formalisms for Ontologies and Concept Spaces}
For a systematic positioning of our approach we use of the triptych allegory proposed by Mayr and Thalheim \cite{thalheim20triptych}. They define conceptual modeling as tripartite consisting of an \textit{encyclopedic dimension} for grounding the semantics in an ontology or concept space, a \textit{language dimension} for the definition of language terms and valid expressions, and the \textit{conceptual modeling dimension} in between as a link between term and concept space.
We are mainly interested in a formalization of the language dimension and acknowledge that in the encyclopedic dimension there also exist various attempts to formalization, like the KL-ONE family \cite{BRACHMAN1989klone} and Description Logic \cite{baader07descriptionlogic}.
Also the formal system of a conceptualization of domains as basis for truthful modeling languages proposed by Guizzardi et al. \cite{guizzardi07,guizzardi15types} has to be located in the encyclopedic dimension and has therefore to be distinguished from our goal. 
In this theory of ontologically-driven conceptual modeling fruitful for the objective of a domain-faithful grounding for modeling languages, the language dimension is an a-posteriori concept implicitly obtained from ontological considerations. 

\subsection{Formalisms for Languages}
When focusing on formalizations in the language dimension the existing approaches can be categorized according to the underlying theory they use, which is mostly graph theory, set theory or logic. All three of them offer concepts for the concrete structural behav\-iour of the elements to be described. In the following we 
present examples illustrating the shortcomings of the former two and argue why logic provides the most canonical approach.

In the domain specific language KM3 presented by Jouault and Bezivin models are defined as directed\linebreak multi-graphs conforming to another model, the metamodel, itself a graph \cite{jouault2006km3}. Using this formalism the authors define a self-descriptive metametamodel and deduce a domain specific language to specify metamodels. This approach puts an emphasis on the graph-like box-and-line structure of models, rather than on the linguistic aspects and similar to MOF has a narrow focus on software structure specification.

A system based on set-theory is the formalization of Ecore and EMOF proposed by Burger \cite[2.3.2]{burger14mofsettheory} which uses the formal description of concepts from the OCL specification \cite[A.1]{OMGGroup2014}. Set theory comprises very basic concepts describing structures, only admiting the subsumption of elements in sets and set hierarchies. It holds no further information about the semantics of the elements.

Also the FDMM formalism introduced by Fill et al. addressing conceptual modeling domains in a broader variety uses set theory to specify metamodels and models \cite{fill12fdmm}. The authors explicitly aim at a formalization of metamodels realized with the metamodeling platform ADOxx \cite{adoxx1} and do not claim to be applicable for platform-independent specifications.

Neither graph theory, basis of KM3, nor set theory, basis of FDMM and the MOF formalization by Burger,  do justice to the linguistic character of modeling languages and provide canonical concepts for the definition of a set of terms and for instantiation, an essential characteristic of modeling languages. 
Therefore the technique and semantics of this instantiation relation between model and metamodel has to be constructed ad-hoc and lacks the beneficial knowledge stack of established theories dealing with linguistic structures.

\subsection{Formalisms Based on Logic}
Formal languages as defined in mathematical logic inherently comprise the concept of instantiation as interpretation of the signature in logical terms, and they provide a rich knowledge base about their properties.
Therefore, in current research the notion of modeling languages as formal languages in the sense of mathematical logic is receiving increasing attention \cite{delcambre18referenceframework,guarino19,olive07conceptual,Partridge13,thalheim11chapter}. 

In their investigation of formal foundations of \linebreak domain-specific languages Jackson and Sztipanovits introduce typed predicate logic to handle object types in models \cite{Jackson:2006:TFF:1176887.1176896,jackson09formalsemantics}. 
They indeed treat modeling languages as formal languages but they do not adopt the concept of a language interpretation, i.e. instantiation for model instances, but rather consider a model to be a set of valid statements about the model. 
This is also true for Telos \cite{koubarakis2020telosretrospective}, which builds on the premise that the concepts of entities and relations are omitted and replaced by propositions constituting the knowledge base. The choice of typed first-order logic for the formalization of these propositions is natural and explained in great detail in \cite{koubarakis89telosformal}. Similar to Jackson and Sztipanovits knowledge is represented solely as a set of sentences in the formal language.
In our approach on the other hand we do not adopt the transformation of models into propositions but rather directly deal with the ubiquitous concepts of objects and relations and an instantiation hierarchy between models and metamodels. 
This leads to a different view on models. In the attempts above a model is constituted by statements, whereas in our approach these statements are used as constraints restricting valid expressions using the proposed signature.

In his work on the theory of conceptual models, Thalheim describes modeling languages as being based on a signature $\Sigma$ comprising a set of postulates, i.e. sentences expressed with elements of $\Sigma$ \cite{thalheim11chapter}. Models are defined as language structures satisfying the postulates, which canonically corresponds to the concept of instantiation of a metamodel. 
We go one step further and concretely point out how to capture the core concepts of a modeling language in a signature $\Sigma$ to unify the method of formalizing a language. This then enables us to investigate the class of formal modeling languages, compare formalized languages, reuse components and develop generic methods for language fusion, model transformation etc. independent of a concrete language.

In summary, the literature review suggests that the 
structure of modeling languages including its linguistic character can be grounded in the concepts of formal languages.  
Therefore, in the work at hand we propose a formal definition of modeling languages in which we concretely specify the modeling concepts and their formal equivalent in logical terms with the prospect of successive elaboration. 

\section{Definition of Formal Modeling Languages} \label{secformalmodelinglanguages}

The intended definition shall serve as a cornerstone for a common way of formalizing modeling languages, which thereby become comparable, reusable and modularizable. 
A formal definition for modeling languages in general enables an investigation of common features of the resulting subclass of formal languages as well as a sound mathematical foundation for their functionality.
We build on a survey conducted by Kern et al. \cite{kern11metameta} on common concepts in the meta$^2$models of six established metamodeling platforms. The definition below incorporates all concepts identified in at least half of the investigated platforms. These are object types, relation types (binary), attributes (multi-value), inheritance (for object types, single), and a constraint language. Other concepts identified in \cite{kern11metameta} which are not yet included are roles, ports, inheritance of relations, n-ary relations, and models in the sense of model types.

These concepts mainly coincide with the core concepts introduced for conceptual modeling of information systems by Olivé \cite{olive07conceptual}. Additional concepts mentioned in Olivé's work which are of high interest but not yet included in our approach are derived types, generic relation types, and powertypes.

\subsection{A Definition based on Predicate Logic}
We use typed (also called sorted) predicate logic in this approach. The mathematical basics can be found in textbooks on logic or mathematics for computer science, e.g. \cite{enderton01logic,mazzola06lmathematics}. 
Some remarks on notation: To ease the differentiation between language and model level, we use capital letters for the symbols of the language and lowercase letters for the elements of the model.
\begin{definition} \label{defmodlang}
	A (formal) modeling language $\mathcal{L}$ consists of a typed signature $\Sigma=\{\mathcal{S}, \mathcal{F}, \mathcal{R}, \mathcal{C}\}$ and a set \textbf{C} of sentences in $\mathcal{L}$ for the constraints, where:
	\begin{itemize}
		\item $\mathcal{S}$ is a set of types, which can be further divided into three disjoint subsets $\mathcal{S}_O$, $\mathcal{S}_R$, and $\mathcal{S}_D$ for object types, relation types and data types;
		\begin{itemize}
			\item the type set $\mathcal{S}_O$ is strictly partially ordered with order relation $<_O \subset \mathcal{S}_O \times \mathcal{S}_O$ to indicate the inheritance relation between the corresponding object types;
			\item the type set $\mathcal{S}_D$ can contain simple types $T$ for value domains of single value attributes, or product types $T'= T_1\times T_2 \times \cdots \times T_n$ for value domains of n-ary multi-value attributes $(n>1)$, where the i-th value is of type $T_i \in \mathcal{S}_D \cup \mathcal{S}_O \cup \mathcal{S}_R $;
		\end{itemize}
		\item $\mathcal{F}$ is a set of typed function symbols such that:
		\begin{itemize}
			\item for each relation type \textbf{R} in $\mathcal{S}_R$ there exist two function symbols $F_s^{\textbf{R}}$ and $F_t^{\textbf{R}}$ with domain type \textbf{R} $\in \mathcal{S}_R$ and codomain type $\textbf{O}_s, \textbf{O}_t \in \mathcal{S}_O$ assigning the source and target object types to a relation;
			\item for each single-value attribute $\textbf{A}$ of an object or relation type $\textbf{T}$ there exists a function symbol $F^\textbf{A}$ with domain type $\textbf{T}$ and codomain type an element in $ \mathcal{S}_D \cup \mathcal{S}_O \cup \mathcal{S}_R $ assigning the simple data type or referenced object type or relation type to the attribute; 
			\item for each multi-value attribute $\textbf{A}$ of an object or relation type  $\textbf{T}$ there exists a function symbol $F^\textbf{A}$ with domain type $\textbf{T}$ and codomain type an product type in $ \mathcal{S}_D $;
		\end{itemize}
		\item $\mathcal{R}$ is a set of typed relation symbols containing $<_O$;
		\item $\mathcal{C}$ is a set of typed constants to specify the possible values $c_i$ of a simple type $\textbf{T}\in \mathcal{S}_D$ of the attributes;
		\item the set \textbf{C} is a set of sentences in $\mathcal{L}$ constraining the possible models, also called the postulates of the language.
	\end{itemize}
\end{definition}
This definition explicates the formalization of the essential modeling concepts of a language, i.e. object types and inheritance, binary directed relation types and \linebreak single- or multi-value attributes. 
Note that the definition does not prohibit the existence of additional symbols in the signature, so broader concepts like n-ary relations can optionally be included and are topic of further investigation. 
Also structures beyond the visual elements of a model can be included, e.g. paths as transitive relations or substructures comprising several elements.

We want to point out, that relation types are defined on the same level as object types, not subordinate to them. 
This highlights their significance for a model beyond mere arrows and allows for 
defining attributes of relations, multiple relations of the same type between the same two objects, as well as for inheritance of relation types.

With the data types and constants we can define attribute domains like integers via specifying a type called $\mathbb{N}$ and constant symbols 0,1,2,3, ... in $\mathcal{C}$ of type $\mathbb{N}$ for the numbers, or enumeration lists like a person's gender via specifying a type called \textit{gender} and constant symbols \textit{male, female}, and \textit{else} in $\mathcal{C}$. The elements of the simple or product types of $\mathcal{S}_D$ are typically not visible in graphical models. They are exclusively used for specifying attribute domains.

Note that if we assume the set of constants for attribute domains to be finite, models are always finite, because by construction they contain only finitely many objects and relations.

\begin{definition} \label{defmodel}
	A model $\mathcal{M}$ of a language $\mathcal{L}$ with typed signature $\Sigma=\{\mathcal{S}, \mathcal{F}, \mathcal{R}, \mathcal{C}\}$ is an $\mathcal{L}$-structure conforming to the language constraints \textbf{C}, i.e. $\mathcal{M}$ consists of 
	\begin{itemize}
		\item a universe $\mathcal{U}$ of typed elements respecting the type hierarchy, that is 
		\begin{itemize}
			\item for each $\textbf{T}$ in $\mathcal{S}$ there exists a set $\mathcal{U}_\textbf{T} \subset \mathcal{U}$ and $\mathcal{U} = \bigcup_{\textbf{T}\in \mathcal{S}} \mathcal{U}_\textbf{T}$;
			\item all sets $\mathcal{U}_{\textbf{T}}$ for $T\in \mathcal{S}_O\cup \mathcal{S}_R$ have to be pairwise disjoint except for sets $\mathcal{U}_{\textbf{O}_1}$ and $\mathcal{U}_{\textbf{O}_2}$ with $\textbf{O}_1, \textbf{O}_2 \in \mathcal{S}_O$ where $\textbf{O}_1 <_O \textbf{O}_2$. In this case $\mathcal{U}_{\textbf{O}_1}$ must be a subset of $\mathcal{U}_{\textbf{O}_2}$, i.e. $\mathcal{U}_{\textbf{O}_1}\subseteq \mathcal{U}_{\textbf{O}_2}$;
		\end{itemize}
		\item an interpretation of the function symbols in $\mathcal{L}$, i.e. for each function symbol $F\in \mathcal{F}$ with domain type $\textbf{T}_1 \times \ldots \times \textbf{T}_n$ and codomain type $\textbf{T}$ a function $f:\mathcal{U}_{\textbf{T}_1}\times \ldots \times \mathcal{U}_{\textbf{T}_n} \rightarrow \mathcal{U}_\textbf{T}$;
		\item an interpretation of the relation symbols in $\mathcal{L}$, i.e. for each relation symbol $R\in \mathcal{R}$ with domain type $\textbf{T}_1 \times \ldots \times \textbf{T}_m$ a relation $r \subset \mathcal{U}_{\textbf{T}_1}\times \ldots \times \mathcal{U}_{\textbf{T}_m}$;
		\item for each simple type $\textbf{T}\in \mathcal{S}_D$ and constant $C \in\mathcal{C}$ of type $\textbf{T}$ an interpretation $c\in\mathcal{U}_{\textbf{T}}$;
		\item for each constraint $\phi$ in \textbf{C} the model $\mathcal{M}$ satisfies $\phi$, i.e. $\mathcal{M}\models \phi$.
	\end{itemize}
\end{definition}
This definition of models as language structures goes beyond a visualisation and considers models as knowledge structures as described in \cite{buchmann2019educationdesignproblem}. Thereby we overcome several shortcomings of graphical representations, like the missing depiction of attributes and their domains in models or the visual mixing of the metarelation inheritance with the definition of relation types in metamodels.

\subsection{Running Example Petri Nets}
\begin{figure}
	\centering
	\includegraphics[width=0.7\linewidth]{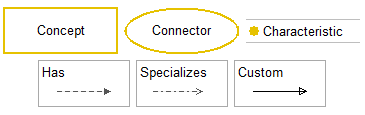}
	\caption{Notation excerpt of the CoChaCo Method \cite{karagiannis2019cochaco}} \label{figlegend}
\end{figure}
\begin{figure}
	\centering
	\includegraphics[width=\linewidth]{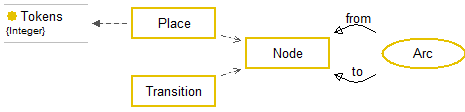}
	\caption{A metamodel of Petri Nets} \label{petrinetMM}
\end{figure}

We will now illustrate the definition on the example of the Petri Net modeling language.
For the visualization of the metamodel we use the notation of CoChaCo, a method to support the creative process of modeling method design \cite{karagiannis2019cochaco}. This method comprises concrete syntax for most of the concepts contained in Definition \ref{defmodlang} with a slightly different naming, see Figure \ref{figlegend}. 
\begin{example} \label{expnmm}
	The Petri Net Modeling Language $\mathcal{PN}$\\
	The Petri Net metamodel depicted in Figure \ref{petrinetMM} comprises three object types \textbf{Node}  (\textbf{No}), \textbf{Place}  (\textbf{Pl}), and \textbf{Transition}  (\textbf{Tr}) constituting $\mathcal{S}_O$. Thereby \textbf{Place} and \textbf{Transition} inherit from \textbf{Node}, i.e. \textbf{Place} $<_O$ \textbf{Node} and \textbf{Transition} $<_O$ \textbf{Node}. Furthermore, the language comprises only one relation type \textbf{Arc} element of $\mathcal{S}_R$. For the attribute \textbf{Tokens} of object type \textbf{Place} we need a type $\mathbb{N}$ with the usual addition $+$ and order relation $<_{\mathbb{N}}$ as well as constants in $\mathcal{C}=\{0,1,2,...\}$ all of type $\mathbb{N}$. The set $\mathcal{S}$ of types is then the union $\mathcal{S}=$ $\mathcal{S}_O \cup \mathcal{S}_R \cup \mathcal{S}_D$=\{\textbf{Node}, \textbf{Place}, \textbf{Transition}, \textbf{Arc}, $\mathbb{N}$\}. For the relation \textbf{Arc} we have to specify the source and target object types by introducing two function symbols $F_s^\text{Arc}$ and $F_t^\text{Arc}$ both with domain \textbf{Arc} and codomain \textbf{Node}. For the attribute \textbf{Tokens} we introduce a function symbol $F^{Tokens}$ with domain \textbf{Place} and codomain $\mathbb{N}$ assigning each place instance a number of tokens. Finally we have to define the constraints of the language. 
	These rules are not contained in a graphical metamodel. 
	In existing specifications they are mainly specified with natural language or OCL. 
	In the predicative formalization at hand constraints are an integral part of the language. Following four sentences written in the alphabet of $\mathcal{PN}$ ensure \textbf{Node} to be abstract, i.e. any element in \textbf{Node} lies either in \textbf{Place} or in \textbf{Transition} (\ref{eqpnlabstract}), as well as the alternation of types of the elements connected by an arc (\ref{eqpnlalt1}, \ref{eqpnlalt2}) and the prohibition of multiple arcs between the same two elements (\ref{eqpnlsinglearc}). For ease of readability we abuse the notation $\forall x \in \textbf{T}$ for $x$ being of type $\mathbf{T}$ instead of using the type specific quantifier $\forall_{_\mathbf{T}}x$.
	\begin{gather}
	\forall x \in \textbf{No}\ \exists y \in \textbf{Pl}, z \in \textbf{Tr}\ (x=y \lor x=z) \label{eqpnlabstract}\\
	\nexists x,y \in \textbf{Pl}, u \in \textbf{Arc}\ (F_s^{Arc}(u) = x \land F_t^{Arc}(u) = y) \label{eqpnlalt1}\\
	\nexists x,y \in \textbf{Tr}, u \in \textbf{Arc}\ (F_s^{Arc}(u) = x \land F_t^{Arc}(u) = y) \label{eqpnlalt2}\\
	\forall u,v \in \textbf{Arc}\ ((F_s^{Arc}(u)=F_s^{Arc}(v)\land \nonumber\\
	\indent F_t^{Arc}(u)=F_t^{Arc}(v)) \implies u=v)\label{eqpnlsinglearc}
	\end{gather}
\end{example}
\begin{figure}
	\centering
	\includegraphics[width=\linewidth]{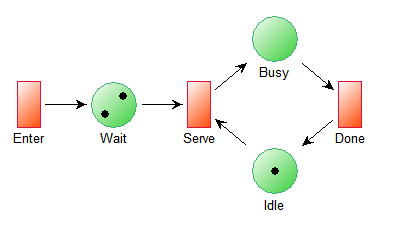}
	\caption{A Petri Net model depicting a simple barber shop scenario} \label{petrinet}
\end{figure}
\begin{example} \label{petrinetmodelexample}
	A Petri Net Model\\
	A Petri Net model depicting a simple barber shop scenario is shown in Figure \ref{petrinet}. Its formalization, i.e. the corresponding $\mathcal{PN}$-structure, looks as follows:
	The universe of places $\mathcal{U}_P$ contains three elements $\mathcal{U}_P$= $\{\textbf{w}(ait)$, $\textbf{b}(usy)$, $\textbf{i}(dle)\}$. The universe of transitions $\mathcal{U}_T$ comprises three elements $\mathcal{U}_T$=$\{\textbf{e}(nter)$, $\textbf{s}(erve)$, $\textbf{d}(one)\}$. Six arc elements exist in $\mathcal{U}_A= \{\textbf{a}_1, \textbf{a}_2, \textbf{a}_3, \textbf{a}_4, \textbf{a}_5, \textbf{a}_6\}$ with source and target $f_s^{Arc}(\textbf{a}_1)=\textbf{e}$, $f_t^{Arc}(\textbf{a}_1)=\textbf{w}$, $f_s^{Arc}(\textbf{a}_2)=\textbf{w}$, $f_t^{Arc}(\textbf{a}_2)=\textbf{s}$, $f_s^{Arc}(\textbf{a}_3)=\textbf{s}$, $f_t^{Arc}(\textbf{a}_3)=\textbf{b}$, $f_s^{Arc}(\textbf{a}_4)=\textbf{b}$, $f_t^{Arc}(\textbf{a}_4)=\textbf{d}$, $f_s^{Arc}(\textbf{a}_5)=\textbf{d}$, \linebreak $f_t^{Arc}(\textbf{a}_5)=\textbf{i}$, $f_s^{Arc}(\textbf{a}_6)=\textbf{i}$, and $f_t^{Arc}(\textbf{a}_6)=\textbf{s}$. For the attribute type and values the natural numbers $\mathbb{N}$ are included in the model, $\mathcal{U}_{\mathbb{N}}=\{0,1,2,...\}$. The instantiation of the attribute \textit{Tokens} looks as follows: $f^{Tokens}(\textbf{w})=2$, $f^{Tokens}(\textbf{b})=0$ and $f^{Tokens}(\textbf{i})=1$. 
	We can easily check that the formalized model satisfies all postulates \ref{eqpnlabstract}--\ref{eqpnlsinglearc} of the language $\mathcal{PN}$.
\end{example}

Notice that the formalized model in Example \ref{petrinetmodelexample} and the graphical model in Figure \ref{petrinet} represent the same thing. They are merely alternative ways of describing a system but with different merits. Whereas the graphical model is easy and fast to comprehend, the formal model is precise and complete, as attribute values are often not legible from a pictoral model. This can be compared to the different representation forms of a graph -- once as a graphical depiction and once as an adjacent matrix.

\section{\texttt{M2FOL} -- MetaModel 2 First Order Logic\\A Formal Modeling Language for Metamodels} \label{secm2fol}

\begin{figure}
	\centering
	\includegraphics[width=0.8\linewidth]{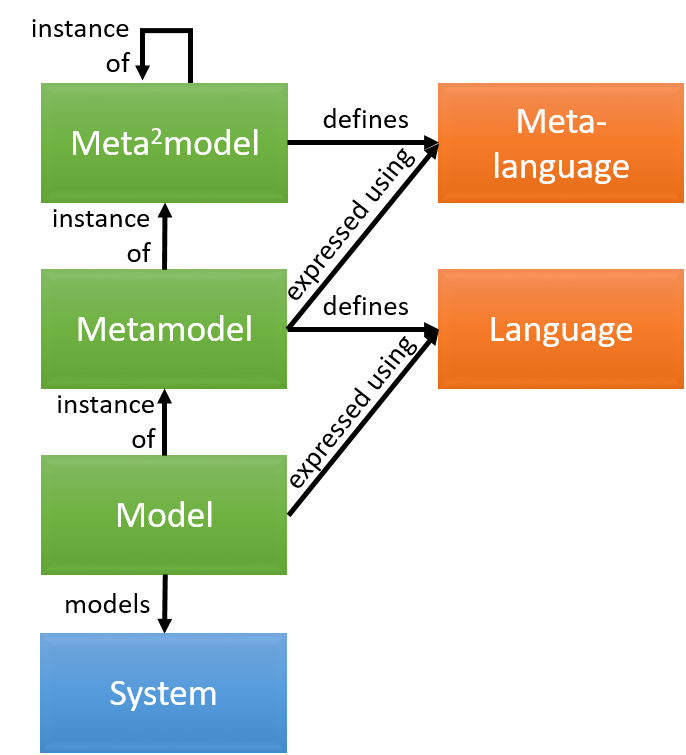}
	\caption{The four-layer metamodeling stack based on \cite{kuhne2006matters}} \label{figmodelingstack}
\end{figure}

Metamodels are models themselves expressed in a metamodeling language. We propose a formal modeling language in the sense of Definition \ref{defmodlang} for metamodels called \texttt{M2FOL}, i.e. a metamodeling language to be exact. This language is capable of describing precisely the concepts explicated in Definition \ref{defmodlang}.
In general meta$^2$models of metamodeling languages are supposed to be self-de\-scribing, which results in a four-layer metamodeling stack as depicted in Figure \ref{figmodelingstack}. We will show, that the metamodel of \texttt{M2FOL}, a meta$^2$model by nature, also partakes of this property.

\subsection{Definition of \texttt{M2FOL}}
We stick to the notational convention of capital letters for elements on the language level and lowercase letters for elements on the model level. To indicate the metalevel of \texttt{M2FOL} and metamodels we use the typewriter font for meta symbols and elements. For ease of readability we write $F:X\rightarrow Y$ when F is a function with domain type $X$ and codomain type $Y$. Nevertheless, the instantiation is then a function $f:\mathcal{U}_X\rightarrow \mathcal{U}_Y$ defined on universes of typed elements.
To be consistent in the naming of the symbols in \texttt{M2FOL} we distinguish between an attribute type on meta level and an attribute as the concrete assignment of a value to an element on model level.

With \texttt{M2FOL} we want to model \textbf{object types} and \textit{inheritance} relations between them, \textbf{relation types} connected to their \textit{from} and \textit{to} object types, \textbf{attribute types} and their \textbf{data types}, and possible \textbf{data}.
According to Definition \ref{defmodlang} all the bold concepts constitute a type in $\mathcal{S}_O$ in \texttt{M2FOL}, whereas all italic concepts make up a type in $\mathcal{S}_R$ in \texttt{M2FOL}. The types \textit{inheritance}, \textit{from}, and \textit{to}, furthermore require assignment functions for source and target specification. 
Data types and data are necessary for defining attribute domains and its values, e.g. the domain $\mathbb{N}_{0..10}$ and values $\{0,1,2,...,9,10\}$ or an enumeration list domain \textit{gender} with values \textit{male}, \textit{female}, \textit{else}. Attribute types need the assignment of owning type and value domain.

\begin{definition} \label{defm2fol}
	The metalanguage \textnormal{\texttt{M2FOL}} is a modeling language with signature $\Sigma=\{\mathcal{S}, \mathcal{F}, \mathcal{R}, \mathcal{C}\}$ with the set of types split in $\mathcal{S}=\mathcal{S}_O \cup \mathcal{S}_R \cup \mathcal{S}_D$, where:
	\begin{itemize}
		\item $\mathcal{S}_O$ consists of the types \texttt{O}(bject) \texttt{T}(ype), \texttt{R}(elation) \texttt{T}(ype), \texttt{A}(ttribute) \texttt{T}(ype), \texttt{D}(ata) \texttt{T}(ype), and \texttt{D}(ata), furthermore two supertypes: \texttt{ORT}(ype), and \linebreak\texttt{DORT}(ype) $ \mathcal{S}_O = \{\texttt{OT}, \texttt{RT}, \texttt{AT}, \texttt{DT}, \texttt{D}, \texttt{ORT}, \texttt{DORT}\}$;
		\begin{itemize}
			\item The types \texttt{OT}, and \texttt{RT}, inherit from \texttt{ORT}, the types \texttt{ORT}, and \texttt{DT} inherit from \texttt{DORT}:  $\texttt{OT}<_O \texttt{ORT}$, $\texttt{RT}<_O \texttt{ORT}$, $\texttt{ORT} <_O \texttt{DORT}$, $\texttt{DT}<_O \texttt{DORT}$;
		\end{itemize}
		\item $\mathcal{S}_R$ consists of the types \texttt{Inh}(eritance), 
		\texttt{Fr}(om), and \texttt{To}: 
		$\mathcal{S}_R = \{\texttt{Inh}, 
		\texttt{Fr}, \texttt{To}\}$;
		\item $\mathcal{S}_D$ contains product types $\texttt{DORT}^n$ for all $n>1$ as well as a type $\texttt{T}_{\texttt{DORT}}$ for the union of all $\texttt{DORT}^n:\texttt{T}_{\texttt{DORT}}=\bigcup_i \texttt{DORT}^i$
		\item the set of function symbols consists of following elements: 
		\begin{itemize}
			\item two symbols $\texttt{F}_{s}^{\texttt{Inh}}$ and $\texttt{F}_{t}^{\texttt{Inh}}$ assigning source and target to \texttt{Inh}-typed relations: $\texttt{F}_{s}^{\texttt{Inh}}:\texttt{Inh}\rightarrow \texttt{OT}$, $\texttt{F}_{t}^{\texttt{Inh}}:\texttt{Inh}\rightarrow \texttt{OT}$;
			\item two symbols $\texttt{F}_{s}^{\texttt{Fr}}$ and $\texttt{F}_{t}^{\texttt{Fr}}$ assigning source and target to \texttt{Fr}-typed relations: $\texttt{F}_{s}^{\texttt{Fr}}:\texttt{Fr}\rightarrow \texttt{RT},$ $\texttt{F}_{t}^{\texttt{Fr}}:\texttt{Fr}\rightarrow \texttt{OT}$;
			\item two symbols $\texttt{F}_{s}^{\texttt{To}}$ and $\texttt{F}_{t}^{\texttt{To}}$ assigning source and target to \texttt{To}-typed relations: $\texttt{F}_{s}^{\texttt{To}}:\texttt{To}\rightarrow \texttt{RT},\ \texttt{F}_{t}^{\texttt{To}}:\texttt{To}\rightarrow \texttt{OT}$;
			\item two symbols $\texttt{F}_{val}$ and $\texttt{F}_{type}$ assigning to an attribute type its value domain and the object or relation type it belongs to. The value assignment can be a reference or a n-valued type in $\texttt{DORT}^n$: $\texttt{F}_{val}: \texttt{AT} \rightarrow \bigcup_i (\texttt{DORT})^i,\ \texttt{F}_{type}:\texttt{AT} \rightarrow \texttt{ORT} $;
			\item a symbol $\texttt{F}_{\texttt{DT}}$ to assign a data type to a data element: $\texttt{F}_{\texttt{DT}}:\texttt{D}\rightarrow \texttt{DT}$;
		\end{itemize}
		\item $\mathcal{R}$ consists of a symbol $<_{\texttt{OT}}$ transitively extending the inheritance relation given by \texttt{Inh} to a strict partial order on the set of object types 
		$\mathcal{R} = \{ <_{\texttt{OT}}\ \subset \texttt{OT}\times \texttt{OT}
		\}$.
	\end{itemize}
	
	The postulates of the language (for brevity we use the abbreviation $xry$ for relation $r$ of type $\textbf{T}$, $x$ being $\texttt{F}_s^{\textbf{T}}(r)$ and $y$ being $\texttt{F}_t^{\textbf{T}}(r)$):
	\begin{gather}
		\forall x,y,z \in \texttt{OT}, v,w \in \texttt{Inh}\ (xvy, xwz \implies \nonumber \\
		\indent y=z \land v=w)\label{eq2}\\
		\forall x,y \in \texttt{OT}, u \in {\texttt{Inh}}\ (xuy \implies x<_\texttt{OT} y)\label{eqinhleq1}\\
		\forall x, y \in  \texttt{OT}\ \exists z \in  \texttt{OT}, u \in {\texttt{Inh}}\ (x<_{\texttt{OT}}y \implies \nonumber\\
		\indent xuy \lor (xuz \land z<_{\texttt{OT}}y)) \label{eqinhleq2}\\
		\forall x\in \texttt{RT}\ \exists y,z \in \texttt{OT}, u \in \texttt{Fr}, v \in \texttt{To}\ (xuy \land xvz)\label{eqrelstart}\\
		\nexists u,v \in {\texttt{Fr}}\ (\texttt{F}_s^{\texttt{Fr}}(u)=\texttt{F}_s^{\texttt{Fr}}(v) \land u\not = v )\\
		\nexists u,v \in {\texttt{To}}\ (\texttt{F}_s^{\texttt{To}}(u)=\texttt{F}_s^{\texttt{To}}(v) \land u\not = v )\label{eqlast}\\
		\forall x \in \texttt{ORT}\ \exists y \in \texttt{OT}, z \in \texttt{RT} (x=y \lor x=z) \label{eqortabstract}\\
		\forall x \in \texttt{DORT}\ \exists y \in \texttt{ORT}, z \in \texttt{DT} (x=y \lor x=z) \label{eqdortabstract}
	\end{gather}
	For $<_{\texttt{OT}}$ we furthermore require to be a strict partial order, i.e. $<_{\texttt{OT}}$ is transitive, irreflexive and antisymmetric.
\end{definition}
The constraints ensure single inheritance (\ref{eq2}), $<_{\texttt{OT}}$ being the transitive closure of \texttt{Inh}  
under the assumption that all universes are finite (\ref{eqinhleq1}--\ref{eqinhleq2}), the existence and uniqueness of to and from objects of a relation (\ref{eqrelstart}-\ref{eqlast}), and the abstractness of the types \texttt{ORT} and \texttt{DORT} (\ref{eqortabstract}--\ref{eqdortabstract}). 
The absence of cyclic inheritance and self-inheritance follow from the properties of $<_{\texttt{OT}}$.

\subsection{Running Example Petri Nets}
With this language we now can transfer the graphical metamodel of Figure \ref{petrinetMM} to a formal \texttt{M2FOL}-model.
\begin{example} \label{pnmodelformalized}
	The Petri Net Metamodel $\texttt{M}_{NP}$\\
	The universe of object types $\mathcal{U}_\texttt{OT}$ comprises three elements \texttt{n}(ode), \texttt{p}(lace), and \texttt{tr}(ansition). The universe of relation types $\mathcal{U}_\texttt{RT}$ contains one element \texttt{a}(rc). One element \texttt{tok}(ens) is contained in the universe of attribute types $\mathcal{U}_{\texttt{AT}}$. The universe $\mathcal{U}_\texttt{Inh}$ contains the instantiation relations $\texttt{p\_n}$ between \texttt{p} and \texttt{n} as well as $\texttt{tr\_n}$ between \texttt{tr} and \texttt{n}. $\mathcal{U}_{\texttt{Fr}}$ contains the relation $\texttt{a\_from}$ of the source element assignment to the relation type \texttt{a}. $\mathcal{U}_{\texttt{To}}$ contains the relation $\texttt{a\_to}$ of the target element assignment to the relation type \texttt{a}. 
	For these four elements the corresponding source and target elements have to be assigned: 		$\texttt{f}_s^{\texttt{Inh}}(\texttt{p\_n})=\texttt{p}$, $\texttt{f}_t^{\texttt{Inh}}(\texttt{p\_n})=\texttt{n}$, $\texttt{f}_s^{\texttt{Inh}}(\texttt{tr\_n})=\texttt{tr}$, $\texttt{f}_t^{\texttt{Inh}}(\texttt{tr\_n})=\texttt{n}$,
	$\texttt{f}_s^{\texttt{Fr}}(\texttt{a\_from})=\texttt{a}$, $\texttt{f}_t^{\texttt{Fr}}(\texttt{a\_from})=\texttt{n}$,  $\texttt{f}_s^{\texttt{To}}(\texttt{a\_to})=\texttt{a}$, $\texttt{f}_t^{\texttt{To}}(\texttt{a\_to})=\texttt{n}$. 
	From \texttt{Inh} the transitive order relation $<_{\texttt{OT}}$ is deduced: $<_{\texttt{OT}} = \{(\texttt{p},\texttt{n}), (\texttt{tr},\texttt{n})\}$.
	Furthermore there are data values $\{0,1,2,...\}$ in $\mathcal{U}_\texttt{D}$ all of type $\mathbb{N} \in \mathcal{U}_{\texttt{DT}}$, $\texttt{f}_{DT}(i)=\mathbb{N}\ \forall i \in \mathcal{U}_\texttt{D}$. These are needed for the value domain of the attribute type \texttt{tok}, an attribute assigned to \texttt{p}: $\texttt{f}_{type}(\texttt{tok})=\texttt{p}$,  $\texttt{f}_{val}(\texttt{tok})=\mathbb{N}$. 
	In short this can be written as follows:
	\begin{gather}
	\mathcal{U}_{\texttt{OT}}=\{\texttt{n}(ode),\ \texttt{p}(lace),\ \texttt{tr}(ansition)\},\\
	\mathcal{U}_{\texttt{RT}}=\{\texttt{a}(rc)\},\ \mathcal{U}_{\texttt{AT}}=\{\texttt{tok}(en)\},\\  
	\mathcal{U}_{\texttt{Inh}}=\{\texttt{p\_n}, \texttt{tr\_n}\},\ \mathcal{U}_{\texttt{Fr}}=\{\texttt{a\_from}\},\\
	\mathcal{U}_{\texttt{To}}=\{\texttt{a\_to}\},\  
	\mathcal{U}_{\texttt{DT}}=\{\mathbb{N}\},\ 
	\mathcal{U}_{\texttt{D}}=\{0,1,2,...\},\\
	\mathcal{U}_{\texttt{ORT}}=\{\texttt{n}, \texttt{p}, \texttt{tr}, \texttt{a}\}, \mathcal{U}_{\texttt{DORT}}=\{\texttt{n}, \texttt{p}, \texttt{tr}, \texttt{a}, \mathbb{N}\}\\
	<_{\texttt{OT}} = \{(\texttt{p},\texttt{n}), (\texttt{tr},\texttt{n})\}\\
	\texttt{f}_s^{\texttt{Inh}}(\texttt{p\_n})=\texttt{p}, \texttt{f}_t^{\texttt{Inh}}(\texttt{p\_n})=\texttt{n}, \\ \texttt{f}_s^{\texttt{Inh}}(\texttt{tr\_n})=\texttt{tr},  \texttt{f}_t^{\texttt{Inh}}(\texttt{tr\_n})=\texttt{n}, \\ \texttt{f}_s^{\texttt{Fr}}(\texttt{a\_from})=\texttt{a}, \texttt{f}_t^{\texttt{Fr}}(\texttt{a\_from})=\texttt{n}, \\
	\texttt{f}_s^{\texttt{To}}(\texttt{a\_to})=\texttt{a}, \texttt{f}_t^{\texttt{To}}(\texttt{a\_to})=\texttt{n},\\
	\texttt{f}_{type}(\texttt{tok})=\texttt{p}, \texttt{f}_{val}(\texttt{tok})=\mathbb{N}, \\
	\texttt{f}_{DT}(i)=\mathbb{N}\ \forall i \in \mathcal{U}_\texttt{D}
	\end{gather}
\end{example}
This formal metamodel $\texttt{M}_{NP}$ conforms to all constraints \ref{eq2}-\ref{eqdortabstract} and describes the formal language $\mathcal{NP}$ introduced in Example \ref{expnmm}. Their subordination prompts a generic procedure on how to deduce the latter from the former.
In Table \ref{tablealg} we present this procedure as an algorithm. In the right column the algorithm is exemplified on the metamodel of Petri Nets. Compare the resulting language to Example \ref{expnmm}.
\begin{table*}
	\begin{center}
		\def\arraystretch{1.5}
		\begin{tabular}{ m{1em}|m{22em}|m{12em}|m{14em}}
			&  \textbf{\texttt{M2FOL} (Meta)Model to Language-Signature} & \textbf{Mapping}  & \textbf{Application to the Petri Net Metamodel}\\ \hline
			1.  &Each metamodel element \texttt{o} in the set $ \mathcal{U}_{\texttt{OT}}$ defines an object type \textbf{O} of the language. The inheritance relation $<_{\texttt{OT}} \subset \mathcal{U}_{\texttt{OT}} \times \mathcal{U}_{\texttt{OT}}$ must be adopted to the types. & $\texttt{o}\in \mathcal{U}_{\texttt{OT}} \rightarrowtail \textbf{O}\in \mathcal{S}_O,$ $<_\texttt{OT} \rightarrowtail <_O$ & $\texttt{node} \in \mathcal{U}_{\texttt{OT}} \rightarrowtail  \textbf{Node} \in \mathcal{S}_O,$ $\texttt{place} \in \mathcal{U}_{\texttt{OT}} \rightarrowtail  \textbf{Place} \in \mathcal{S}_O,$ $\texttt{trans} \in \mathcal{U}_{\texttt{OT}} \rightarrowtail  \textbf{Trans} \in \mathcal{S}_O$\\ \hline
			2. & Each metamodel element \texttt{r} in the set $\mathcal{U}_{\texttt{RT}}$ defines a relation type \textbf{R} of the language.  & $\texttt{r}\in \mathcal{U}_{\texttt{RT}} \rightarrowtail \textbf{R}\in \mathcal{S}_R$ & $\texttt{arc}\in \mathcal{U}_{\texttt{RT}} \rightarrowtail \textbf{Arc}\in \mathcal{S}_R$ \\ \hline 
			3. & For each relation type $\texttt{r}\in \mathcal{U}_{\texttt{RT}}$ there exist an element $\texttt{s}$ of type \texttt{From} and an element $\texttt{t}$ of type \texttt{To} and both relation elements have as source element \texttt{r},  $\texttt{f}_s^{\texttt{Fr}}(\texttt{s})=\texttt{r}, \texttt{f}_s^{\texttt{Fr}}(\texttt{t})=\texttt{r}$. The assignment $\texttt{f}_t^{\texttt{To}}(\texttt{s})=\texttt{o}_s$ indicates the source object type of $\textbf{R}$, $\texttt{f}_t^{\texttt{To}}(\texttt{t})=\texttt{o}_t$ indicates the target object type of $\textbf{R}$. & $\texttt{r},\texttt{s}, \texttt{o}_s \rightarrowtail \texttt{F}_s^{\text{R}}:\textbf{R}\rightarrow  \textbf{O}_s$; $\texttt{r},\texttt{t}, \texttt{o}_t \rightarrowtail \texttt{F}_t^{\text{R}}:\textbf{R}\rightarrow  \textbf{O}_t$ & $\texttt{arc},\texttt{f}_t^{\texttt{Fr}}(\texttt{a\_from})=\texttt{n} \rightarrowtail \texttt{F}_s^{\text{Arc}}:\textbf{Arc}\rightarrow  \textbf{Node};$ \hspace{3cm} 
			$\texttt{arc},\texttt{f}_t^{\texttt{To}}(\texttt{a\_to})=\texttt{n} \rightarrowtail \texttt{F}_t^{\text{Arc}}:\textbf{Arc}\rightarrow  \textbf{Node}$ \\ \hline
			4. & Each metamodel element \texttt{dt} in $\mathcal{U}_{\texttt{DT}}$ defines a data type \textbf{DT} of the language. 
			Each metamodel element \texttt{d} in $\mathcal{U}_{\texttt{D}}$ with $\texttt{f}_{\texttt{DT}}(\texttt{d})=\texttt{dt}$ becomes a constant symbol $C_d$ in $\mathcal{C}$ of type \textbf{DT}.&
			$\texttt{dt} \in \mathcal{U}_\texttt{DT}\rightarrowtail \textbf{DT} \in \mathcal{S}_{D}$; \hspace{0.5cm}$\texttt{d} \in \mathcal{U}_\texttt{D}\rightarrowtail C_\texttt{d} \in \mathcal{C}$ & 
			$\mathbb{N} \in \mathcal{U}_\texttt{DT}\rightarrowtail \mathbb{N} \in \mathcal{S}_{DT};$ \hspace{1cm} $\texttt{i} \in \mathcal{U}_\texttt{D}, \texttt{f}_{DT}(i)=\mathbb{N} \rightarrowtail \text{i} \in \mathcal{C}$ of type $\mathbb{N}$\\ \hline
			5. & Each metamodel element \texttt{a} in the set $\mathcal{U}_\texttt{AT}$ defines a function symbol $\texttt{F}^\texttt{a}$ of the language. The object or relation type \texttt{a} belongs to, i.e. the domain of $\texttt{F}^\texttt{a}$, is given by the assignment $\texttt{f}_{type}(\texttt{a})=\texttt{t}_{ty}\in \mathcal{U}_\texttt{OT} \cup \mathcal{U}_\texttt{RT} $, its value range, i.e. codomain, by $\texttt{f}_{val}(\texttt{a})=(\texttt{t}_{v_1}, \ldots, \texttt{t}_{v_n})\in (\mathcal{U}_\texttt{OT} \cup \mathcal{U}_\texttt{RT}  \cup \mathcal{U}_\texttt{DT})^n$ &
			$\texttt{a}, \texttt{t}_{ty}, \texttt{t}_{\bar{v}} \rightarrowtail F^\texttt{a}:T_{ty}\rightarrow T_{\bar{v}}$ & 
			$\texttt{tok}, \texttt{f}_{type}(\texttt{tok})=\texttt{place},$ $\texttt{f}_{val}(\texttt{tok})=\mathbb{N}$ $\rightarrowtail$ $F^{Tokens}: \textbf{Place} \rightarrow \mathbb{N}$ \\ \hline
			6. & The constraints of the language have to be added manually, because this information is not determined by the metamodel. & &
		\end{tabular}
		\caption{Algorithm to deduce a formal modeling language signature from its \texttt{M2FOL} metamodel specification} \label{tablealg}
	\end{center}
\end{table*}

\begin{figure*}
	\centering
	\includegraphics[width=0.7\textwidth]{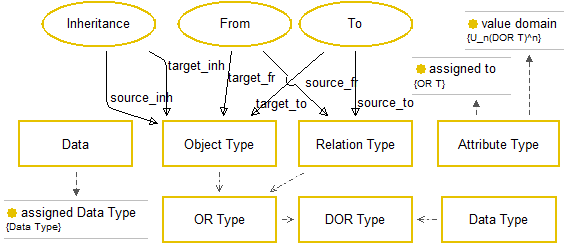}
	\caption{The metamodel of \texttt{M2FOL}} \label{metasquaremodel}
\end{figure*}

\subsection{Meta-perspective on \texttt{M2FOL}}
	Finally we formalize the metamodel of \texttt{M2FOL} as \texttt{M2FOL} model. The graphical metamodel is depicted in Figure \ref{metasquaremodel}.
\begin{example}\label{exm2folmm}Metamodel of \texttt{M2FOL} \\
The \texttt{M2FOL} metamodel contains seven objects of type \texttt{OT}=$\{\texttt{o}(\textit{bject}) \texttt{t}(\textit{ype})$, $\texttt{r}(\textit{elation}) \texttt{t}(\textit{ype})$, $\texttt{a}(\textit{ttribute}) \texttt{t}(\textit{ype})$, $\texttt{d}(\textit{ata}) \texttt{t}(\textit{ype})$, $\texttt{d}(\textit{ata})$, $\texttt{ort}(\textit{ype})$, $\texttt{dort}(\textit{ype})\}$, three objects of type \texttt{RT}=$\{\texttt{inh},$ 
	$\texttt{fr}(\textit{om}),$ $\texttt{to}\}$, three objects of type $\texttt{AT}$= $\{\texttt{val}(\textit{ue})\_\texttt{dom}(\textit{ain})$, $\texttt{ass}(\textit{igned})\_\texttt{to}$, $\texttt{ass}(\textit{igned})\_$ $\texttt{d}(\textit{ata})\texttt{t}(\textit{ype})$,
	many objects of type $\texttt{DT}=\{\texttt{dort}^i \forall i$, \linebreak$\bigcup_i(\texttt{dort})^i\} $ (not visible in the graphical metamodel), 
	four relations of type \texttt{Inh}=$\{\texttt{ot}<\texttt{ort},\texttt{rt}<\texttt{ort},\texttt{ort}<\texttt{dort},\texttt{dt}<\texttt{dort}\} $, three relations of type \texttt{From}=\linebreak$\{\texttt{source\_inh}$, $\texttt{source\_to}$,\- $\texttt{source\_fr}\}$, as well as three relations in \texttt{To}=$\{\texttt{target\_inh}$, $\texttt{target\_to}$, $\texttt{target\_fr}\}$, furthermore 26 assignments of source and target objects, attribute owning types and attribute value types.
	\begin{gather}
	f_s(\texttt{target\_inh})=\texttt{inh}, f_t(\texttt{target\_inh})=\texttt{ot},  \\
	f_s(\texttt{source\_inh})=\texttt{inh}, f_t(\texttt{source\_inh})=\texttt{ot}, \\
	f_s(\texttt{target\_fr})=\texttt{fr}, f_t(\texttt{target\_fr})=\texttt{ot}, \\
	f_s(\texttt{source\_fr})=\texttt{fr}, f_t(\texttt{source\_fr})=\texttt{rt},\\
	f_s(\texttt{target\_to})=\texttt{to}, f_t(\texttt{target\_to})=\texttt{ot}, \\
	f_s(\texttt{source\_to})=\texttt{to}, f_t(\texttt{source\_to})=\texttt{rt}, \\
	f_s(\texttt{ot}<\texttt{ort})=\texttt{ot}, f_t(\texttt{ot}<\texttt{ort})=\texttt{ort}, \\
	f_s(\texttt{rt}<\texttt{ort})=\texttt{rt}, f_t(\texttt{rt}<\texttt{ort})=\texttt{ort}, \\
	f_s(\texttt{ort}<\texttt{dort})=\texttt{ort}, f_t(\texttt{ort}<\texttt{dort})=\texttt{dort}, \\
	f_s(\texttt{dt}<\texttt{dort})=\texttt{dt}, f_t(\texttt{dt}<\texttt{dort})=\texttt{dort}, \\
	f_{type}(\texttt{ass\_dt})=\texttt{d}, f_{val}(\texttt{ass\_dt})=\texttt{dt}\\
	f_{type}(\texttt{ass\_to})=\texttt{at}, f_{val}(\texttt{ass\_to})=\texttt{ort},\\
	f_{type}(\texttt{val\_dom})=\texttt{at}, f_{val}(\texttt{val\_dom})=\bigcup_i(\texttt{dort})^i
	\end{gather}
	On the one hand the construct above is itself a model expressed in the language \texttt{M2FOL}.
	On the other hand this metamodel defines \texttt{M2FOL} as a meta$^2$model. With the algorithm presented above we deduce Definition \ref{defm2fol} from Example \ref{exm2folmm}. So we conclude that the proposed modeling language \texttt{M2FOL} for metamodels is self-\linebreak describing and thereby complete the formalization of the four-layer metamodeling stack.
\end{example}

\begin{figure*}
	\centering
	\includegraphics[width=\textwidth]{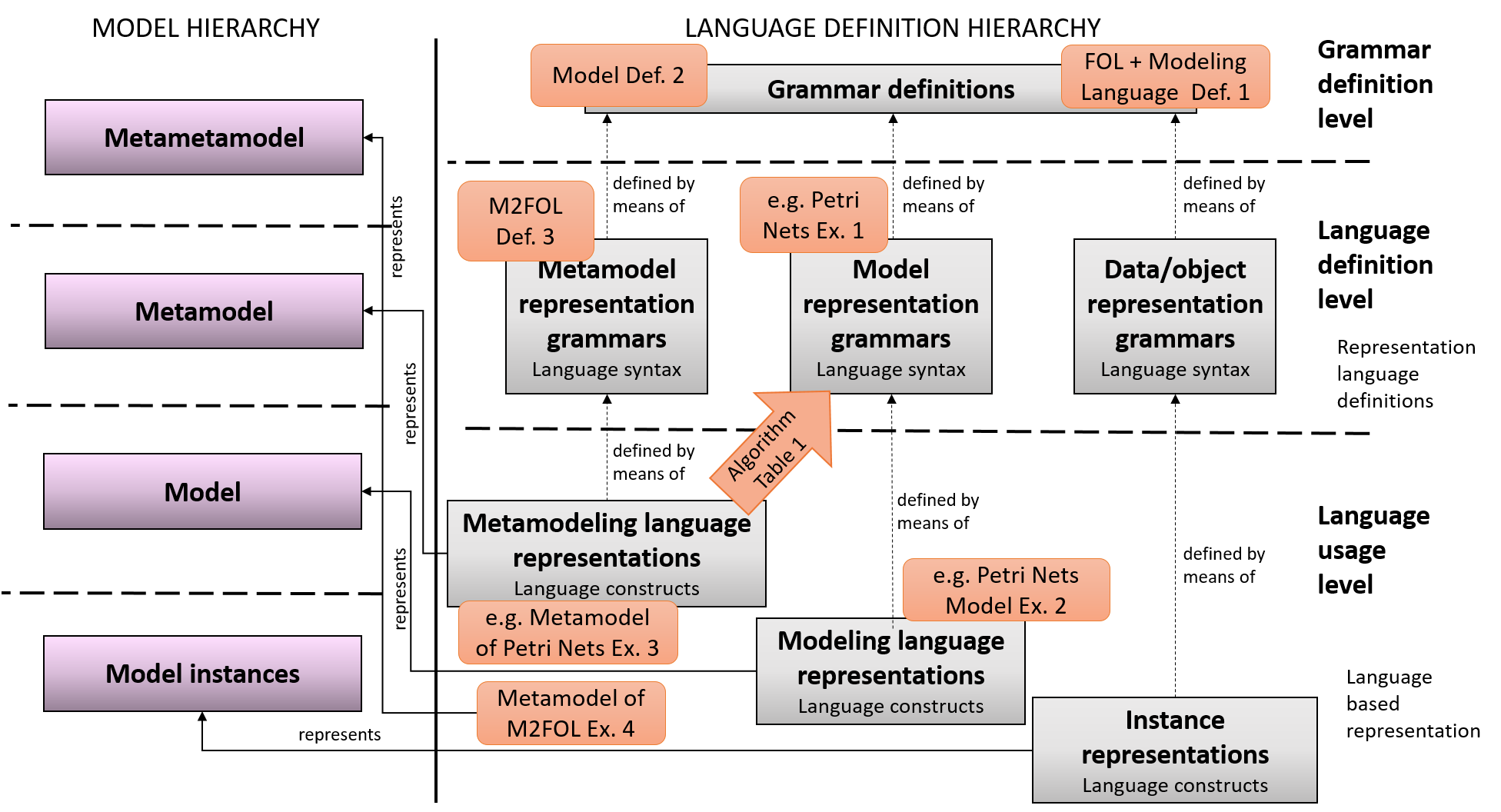}
	\caption{Language Definition and Model Hierarchy adapted from \cite{thalheim20triptych}} \label{figlanguagedefhierarchy}
\end{figure*}

In Figure \ref{figlanguagedefhierarchy} we do a wrap-up of all the presented definitions and examples on the language definition hierarchy proposed by Thalheim and Mayr \cite{thalheim20triptych}. The authors make an explicit distinction between this hierarchy and the model hierarchy. On the \textit{grammar definition level} they allocate the means of defining the language grammars. Examples for elements on this level are EBNF or in our case Definition \ref{defmodlang} and Definition \ref{defmodel} concerning the formal definitions of a modeling language and a model.
The grammars residing on this level are used in the next lower \textit{level of language definition}. Here we can see not only model representation grammars but also metamodel representation grammars. An example for a model representation grammar is the formalized Petri Net language from Example \ref{expnmm}, an example for a metamodel representation grammar is Definition \ref{defm2fol} of \texttt{M2FOL} both using the proposed formalism in Definition \ref{defmodlang} as grammar definition tool.
On the lowest \textit{level of language usage} we find instances of these languages: the Petri Net model from Example \ref{petrinetmodelexample} as a modeling language representation and the metamodel of Petri Net from Example \ref{pnmodelformalized} as a metamodeling language representation defined by means of \texttt{M2FOL}. An example of a meta$^2$modeling language representation on this level is the metamodel of \texttt{M2FOL} also defined by means of \texttt{M2FOL}. This shows that \texttt{M2FOL} is a metamodel representation grammar as well as a meta$^2$model representation grammar.
The algorithm presented in Table \ref{tablealg} allows for the automatic derivation of the language syntax of a model representation grammar on the language definition level from a metamodeling language representation on the language usage level.

\section{Potential and Benefits of Formalized Conceptual Modeling Languages} \label{secoutlookmethods}

In this section we give an outlook to several research topics potentially benefiting from formalizing conceptual modeling languages with the proposed formalism -- these are language interleaving and consistency, operations on models, and translators of platform independent formalizations to platform specific code.
For this purpose, we make use of established concepts from formal language theory.
For reasons of brevity we will not discuss all topics in detail and will restrict to an extensive elaboration only for the first topic of interest:

\subsection{Language Interleaving and Consistency}
Models are means to manage information in highly complex systems in business modeling, in software engineering and many other fields.
The solution to cope with complexity is often seen in the distribution and fragmentation of information between different models or views possibly in different modeling languages, thereby raising the issue of keeping the models consistent \cite{awadid2019consistency,burger16viewbased,burger21vitruvius}.

In the following we demonstrate that language interleaving and the definition of consistency constraints can be easily realized in formalized conceptual modeling languages. 
Depending on the initial situation we can distinguish top-down approaches, where a newly defined or existing language is segmented in several sublanguages or views, and bottom-up approaches, where existing languages are interleaved and their metamodels are amalgamated and equipped with additional constraints \cite{burger19singleunderlyingmodel}. We will discuss the first approach briefly and exemplify the second one in a case study.

\subsubsection{Top-Down Approach}
Expressed in our formalism the top-down approach\linebreak means to restrict the signature $\Sigma$ of a given language $\mathcal{L}$ to subsets $\Sigma_1$ and $\Sigma_2$ of the signature. When working in one view, i.e. with a sublanguage $\mathcal{L}|_{\Sigma_i}$, we are restricted only to the types appearing in $\Sigma_i$. Note that we also have to remove relation types, if their source or target object type was excluded from the signature as well as for attribute types, if their source type or value type was excluded. All constraints considering unavailable types have to be removed. Note also that the signatures $\Sigma_i$ of the sublanguages do not have to be disjoint. While restricting to a sublanguage $\mathcal{L}|_{\Sigma_i}$ and thereby restricting to a concrete view on a system under study we are still interesting in the model as a whole. So we assume that for each view of $\mathcal{L}|_{\Sigma_i}$ there exist correlated models in the other views $\mathcal{L}|_{\Sigma_j}$ being dependent on each other. Therefore we need pairwise constraints between the possible views always considering the signatures of the two relevant languages. If the signatures are not disjoint these constraints contain isomorphisms of elements with a common type to keep the shared structure consistent.

\subsubsection{Bottom-Up Approach}
Expressed in our formalism the bottom-up approach means to fusion the signatures of two given languages $\mathcal{L}_1$ and $\mathcal{L}_2$.
The interleaving of models and their language reaches from simply referencing to elements in other models to a highly dependent content and structure of models in both directions.
There exist different techniques to link conceptual modeling languages \cite{awadid2019consistency}. 
There are also different techniques in the field of logic on how to combine formal languages, e.g. \cite{baader2007,kutz04_econnections}.
The presented attempt is mainly inspired by the former reference.

Uniting two given languages $\mathcal{L}_1$ and $\mathcal{L}_2$ requires uniting their signatures $\Sigma_1$ and $\Sigma_2$.
When doing so, we have to take care for types $T$ occurring in both languages.
To stay compatible with existing models we keep both types and rename them to $T_1$ and $T_2$. 
Furthermore we introduce new function symbols $i:T_1\rightarrow T_2$.
These functions are required to be bijective as we assume, that we want to depict the same situation with both views, i.e. sublanguages. 

With this union the sets of object types and relation types of the new language $\bar{\mathcal{L}}$ are fixed. Also the inheritance relations do not change and stay separated for both initial languages. To create new references and consistency constraints we may introduce new attributes $A$ with $\textbf{F}^A:\textbf{T}_{dom}\rightarrow \textbf{T}_{val}$, where attributed type $\textbf{T}_{dom}$ and value domain $\textbf{T}_{val}$ might stem from different initial languages. To define the attributes and constraints as required we might also have to introduce new product types in $\mathcal{S}_{\textbf{D}}$ and additional function and relation symbols in $\mathcal{F}$ and $\mathcal{R}$ respectively. 

In summary, the newly obtained signature $\bar{\Sigma}$ for the language $\bar{\mathcal{L}}$ looks as follows:
\begin{itemize}
	\item $\bar{\Sigma}=\{\mathcal{S}, \mathcal{F}, \mathcal{R}, \mathcal{C}\}$ with $\bar{\mathcal{S}}=\bar{\mathcal{S}}_O \cup \bar{\mathcal{S}}_R \cup \bar{\mathcal{S}}_D$;
	\item with $\bar{\mathcal{S}}_O=\mathcal{S}_O^1 \cup \mathcal{S}_O^2$ and $\bar{\mathcal{S}}_R=\mathcal{S}_R^1 \cup \mathcal{S}_R^2$;
	\item the set of data types $\bar{\mathcal{S}}_D\supset \mathcal{S}_D^1 \cup \mathcal{S}_D^2$ furthermore contains all newly defined product types $\bar{\textbf{T}} \in (\mathcal{S}_O \cup \mathcal{S}_R \cup \mathcal{S}_D)^i$;
	\item the set $\bar{\mathcal{F}}\supset \mathcal{F}^1 \cup \mathcal{F}^2$ additionally contains the new function symbols;
	\item the set $\bar{\mathcal{R}}\supset \mathcal{R}^1 \cup \mathcal{R}^2$ in addition contains the newly created relation symbols;
\end{itemize}
The constraints \textbf{C} are extended with the requirement that the mappings of the coinciding types $i:T_1\rightarrow T_2$ are bijective, as well as further  defined postulates necessary for information consistency.

\subsubsection{Case Study}
\begin{figure*}
	\centering
	\subfloat[\label{figomodet1}]{\includegraphics[width=0.35\textwidth]{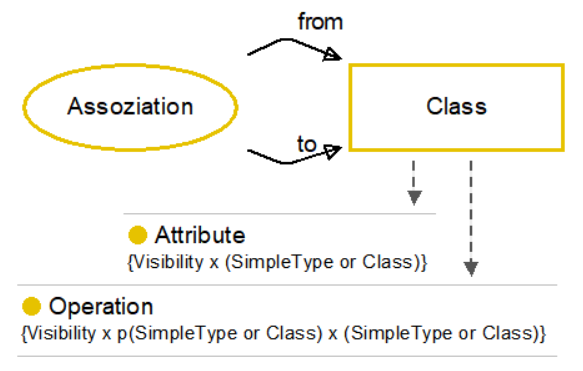}}
	\hfill
	\subfloat[\label{figomodet3}]{\includegraphics[width=0.55\textwidth]{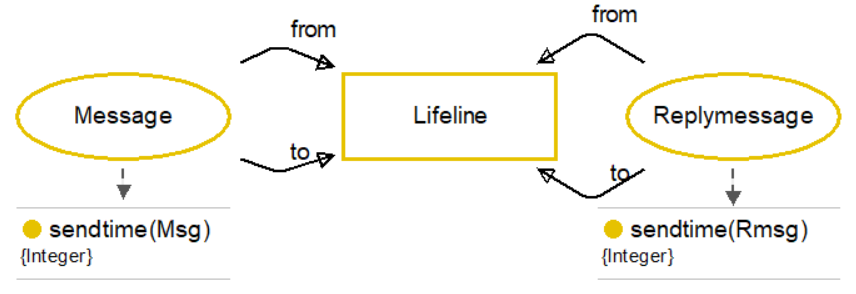}}
	\caption{Simplified metamodels of the UML Class Diagram (a) and Sequence Diagrams (b)} \label{figumlmeta}
\end{figure*}

We will demonstrate the procedure of interleaving on a case study of UML Class Diagrams and Sequence Diagrams. First we have to formalize the initial languages 
$\mathcal{CD}$ of Class Diagrams and $\mathcal{SD}$ of Sequence Diagrams. For an easier comprehension the considered signatures restrict to a subset of the original UML concepts relevant for the connection of the two languages, see Figure \ref{figumlmeta}.

\begin{example} \label{exumlclass}
	The UML Class Diagram Language $\mathcal{CD}$\\
	For our purpose it suffices to consider in the signature $\Sigma_{\mathcal{CD}}$ only one object type \textbf{Class} (\textbf{Cl}) and one relation type \textbf{Association} (\textbf{As}) connecting classes.
	Besides the plain data types \textbf{Visibility} (\textbf{V}) and \textbf{SimpleType} (\textbf{ST}), class diagrams also provide 
	the construct of attributes and operations of classes exhibiting a complex structure themselves.
	We define a data type called \textbf{Attribute} (\textbf{At}) which is a tuple of an element of type \textbf{Visibility} and the value of the attribute, i.e. an object of \textbf{ComplexType} (\textbf{CT}) $=\textbf{SimpleType} \cup \textbf{Class}$. 
	Furthermore we need a data type \textbf{Operation} (\textbf{Op}) which is a tuple of an element of type \textbf{Visibility}, a return type element in \textbf{CT}, and arbitrary many  parameters of type \textbf{CT}. For these parameters we use the union of all product types $(\textbf{CT})^i$. For reasons of brevity we skip the explicit definition of all intermediate product types in the signature and just refer to $\bigcup_i (\textbf{CT})^i$.
	As classes can have several (distinct) attributes and operations, the model attributes \textbf{C}(\textit{lass})\textbf{At}(\textit{tributes}) and \textbf{C}(\textit{lass})\textbf{Op}(\textit{erations}) point to the powersets of types $\wp(\textbf{AT})$ and $\wp(\textbf{OP})$.
	
	The signature of $\mathcal{CD}$ looks as follows:
	\begin{gather}
	\Sigma_{\mathcal{CD}}=\{\mathcal{S}, \mathcal{F}, \mathcal{R}, \mathcal{C}\}, \mathcal{S}=\mathcal{S}_O \cup \mathcal{S}_R \cup \mathcal{S}_D\\
	\mathcal{S}_O=\{\textbf{Class}\}, \mathcal{S}_R=\{\textbf{Association}\}, \\
	\mathcal{S}_D=\{\textbf{Visibility}, \textbf{SimpleType}, \nonumber\\
	\indent \textbf{ComplexType}=\textbf{ST} \cup \textbf{Cl}, \textbf{Attribute}= \textbf{V} \times \textbf{CT}, \nonumber\\
	\indent \textbf{Operation} = \textbf{V} \times\bigcup_i(\textbf{CT})^i\times \textbf{CT}\},\\
	\mathcal{F}=\{F_s^\text{\textbf{As}}:\textbf{As} \rightarrow \textbf{Cl}, F_t^\text{\textbf{As}}:\textbf{As} \rightarrow \textbf{Cl},\nonumber\\
	\indent F^{CAt}:\textbf{Cl}\rightarrow \wp (\textbf{At}),\ F^{COp}:\textbf{Cl}\rightarrow \wp (\textbf{Op}) \} \\
	\mathcal{C}=\{+,-,\sim, String, Integer, Real, Boolean\} 
	\end{gather}
	Thereby, $+, -, \sim$ are of type \textbf{Visibility} and \textit{String, Integer, Real}, and \textit{Boolean} are of type \textbf{SimpleType}.
	
	We do not need any postulates on the language $\mathcal{CD}$.
\end{example}

\begin{example}\label{exumlsequence}
	The UML Sequence Diagram Language $\mathcal{SD}$\\
	Also in this example we restrict to the simplified case of having only one object type \textbf{Lifeline} (\textbf{Ll}) and two relation types \textbf{Message} (\textbf{Msg}) and \textbf{Replymessage} (\textbf{Rmsg}) both connecting lifelines.
	The temporal sequence of messages usually captured in the graphical order of arrows is defined in the attribute \textbf{Sendtime} (\textbf{MSt} and \textbf{RSt}) with value domain $\mathbb{N}$ assigning a point in time to the messages and replymessages.
	To be able to compare sendtimes we need the usual order relation $<_{time} \subset \mathbb{N}\times \mathbb{N}$ and the usual addition function $+_{time}$in the signature:
	\begin{gather}
	\Sigma_{\mathcal{SD}}=\{\mathcal{S}, \mathcal{F}, \mathcal{R}, \mathcal{C}\},\	\mathcal{S}=\mathcal{S}_O \cup \mathcal{S}_R \cup \mathcal{S}_D\\
	\mathcal{S}_O=\{\textbf{Lifeline}\}, \	\mathcal{S}_D=\{\mathbb{N}\},\\
	\mathcal{S}_R=\{\textbf{Message}, \textbf{Replymessage}\}, \\ 
	\mathcal{F}=\{ F^{MSt}:\textbf{Msg}\rightarrow \mathbb{N}, F^{RSt}:\textbf{Rmsg}\rightarrow \mathbb{N}, \nonumber \\
	\indent F_s^\text{{Msg}}:\textbf{Msg} \rightarrow \textbf{Ll}, F_t^\text{{Msg}}:\textbf{Msg} \rightarrow \textbf{Ll}, \nonumber\\
	\indent F_s^\text{Rmsg}: \textbf{Rmsg} \rightarrow \textbf{Ll}, F_t^\text{Rmsg}: \textbf{Rmsg} \rightarrow \textbf{Ll} \nonumber\\
	\indent F_{+_{time}}:\mathbb{N}\times \mathbb{N} \rightarrow \mathbb{N} \}\\
	\mathcal{R}=\{<_{time} \subset \mathbb{N}\times \mathbb{N}\}, \mathcal{C}=\{0,1,2,...\} \text{ of type }\mathbb{N} 
	\end{gather}

	To ensure a reasonable temporal flow of messages we need two language constraints:
	\begin{gather}
	\forall x,y \in \textbf{Msg} (F^{MSt}(x)<_{time}F^{MSt}(y) \lor \nonumber\\
	\indent  F^{MSt}(y)<_{time}F^{MSt}(x)) \lor x=y \label{eqsdsequential}\\
	\forall x \in \textbf{Rmsg}, \exists y \in \textbf{Msg} \nonumber\\
	\indent (F^{MSt}(y)+_{time}1=F^{RSt}(x)) \label{eqsdwait}
	\end{gather}
	Equation \ref{eqsdsequential} restricts diagrams to be sequential, so no two messages are sent at the same time. Equation \ref{eqsdwait} forces the message flow to be synchronous.
	
\end{example} 

\begin{example}
	The Interleaved Modeling Language $\mathcal{CD} \uplus \mathcal{SD}$\\
	In the case of UML class diagrams and sequence diagrams we do not have to take care of identical types.
	We define several new attributes to bind lifelines in a sequence diagram to the classes in the corresponding class diagram:	a reference \textbf{L}(\textit{ife})\textbf{l}(\textit{ine})\textbf{Cl}(\textit{as}), a reference \textbf{Cal}(\textit{led})\textbf{Op}(\textit{eration}) of a message, and a reference \textbf{Re}(\textit{turn})\textbf{Ty}(\textit{pe}) of a replymessage.
	\begin{gather}
	F^{LlCl}: \textbf{Lifeline} \rightarrow \textbf{Class}\\
	F^{CalOp}: \textbf{Message} \rightarrow \textbf{Operation}\\
	F^{ReTy}: \textbf{Replymessage} \rightarrow \textbf{ComplexType}
	\end{gather}
	These references of course require some new constraints.
	To formulate these we need a new relation symbol $\in_{Op}$ and a new function symbol $F_{pr}$ projecting an element of type \textbf{Operation} to its returntype, i.e. the last value of the tuple.
	\begin{gather}
	\in_{Op} \subset \textbf{Operation} \times \wp (\textbf{Operation}),\\
	F_{pr}: \textbf{Operation} \rightarrow \textbf{ComplexType}\\
	\forall x = (x_1,\ldots, x_n) \in \textbf{Operation} \ F_{pr}(x)=x_n
	\end{gather}
	
	With these symbols we can define the additional constraints:
	\begin{gather}
	\forall x \in \textbf{Msg}\nonumber \\ 
	\indent ( F^{CalOp}(x)\in_{Op} F^{COp}(F^{class}(F_t^{Mes}(x)))) \label{eqcalop}\\
	\forall x \in \textbf{Rmsg}\ \exists y \in \textbf{Msg} ( F^{MSt}(y)+1=F^{RSt}(x) \land \nonumber \\ 
	\indent F^{ReTy}(x)=F_{pr}(F^{CalOp}(y))) \label{eqrettype}
	\end{gather}
	Equation \ref{eqcalop} ensures that a message can only call operations of the addressed class. Equation \ref{eqrettype} guarantees that each replymessage follows a message and the returntype is exactly the returntype of the called operation.
\end{example}	
	
The complete language $\mathcal{CD} \uplus \mathcal{SD}$ looks as follows:
\begin{gather}
\Sigma_{\mathcal{CD} \uplus \mathcal{SD}}=\{\mathcal{S}, \mathcal{F}, \mathcal{R}, \mathcal{C}\}, \mathcal{S}_O=\{\textbf{Class}, \textbf{Lifeline}\}, \\
 \mathcal{S}_R = \{\textbf{Association}, \textbf{Message}, \textbf{Replymessage}\}\\
\mathcal{S}_D = \{\textbf{Visibility}, \textbf{SimpleType}, \textbf{ComplexType},\nonumber\\
\indent \textbf{Attribute}, \textbf{Operation}, \mathbb{N}\}\\
\mathcal{F}=\{F_s^\text{As}:\textbf{As} \rightarrow \textbf{Cl}, F_t^\text{As}:\textbf{As} \rightarrow \textbf{Cl},\nonumber\\
\indent F^{CAt}:\textbf{Cl}\rightarrow \wp (\textbf{At}), F^{COp}:\textbf{Cl}\rightarrow \wp (\textbf{Op}), \nonumber\\
\indent F_s^\text{Msg}:\textbf{Msg} \rightarrow \textbf{Ll}, F_t^\text{Msg}:\textbf{Msg} \rightarrow \textbf{Ll},\nonumber\\
\indent F_s^\text{Rmsg}: \textbf{Rmsg} \rightarrow \textbf{Ll}, F_t^\text{Rmsg}: \textbf{Rmsg} \rightarrow \textbf{Ll},\nonumber\\
\indent F^{MSt}:\textbf{Msg}\rightarrow \mathbb{N}, F^{RSt}:\textbf{Rmsg}\rightarrow \mathbb{N},\nonumber\\
\indent	F^{LlCl}: \textbf{Lifeline} \rightarrow \textbf{Class}\nonumber\\
\indent F^{CalOp}: \textbf{Message} \rightarrow \textbf{Operation}\nonumber\\
\indent F^{ReTy}: \textbf{Replymessage} \rightarrow \textbf{ComplexType} \nonumber \\
\indent F_{pr}: \textbf{Operation} \rightarrow \textbf{ComplexType} \nonumber \\
\indent F_{+_{time}}:\mathbb{N}\times \mathbb{N} \rightarrow \mathbb{N}\}\\
\mathcal{R}=\{<_{time} \subset \mathbb{N}\times \mathbb{N}, \in_{Op} \subset \textbf{Op} \times \wp (\textbf{Op})\}\\
\mathcal{C}= \{+,-,\sim, String, Integer, Real, Boolean,\nonumber \\
\indent 0,1,2,...\}
\end{gather}

With the newly generated language each model contains all information of both views, the structural view of class diagrams as well as the procedural view of sequence diagrams. Of course when viewing the model we only consider the model restricted to a sublanguage, $\mathcal{CD}$ or $\mathcal{SD}$ but in the background all elements of both reside in the ``supermodel''. This means all information is captured in the model at all points in time and at the same time kept consistent due to the newly introduced postulates.
This conforms to the idea of the single underlying model as proposed by Burger et al. \cite{burger21vitruvius,burger19singleunderlyingmodel}.\\

\subsection{Operations on Models} \label{subsecop}
Model functionality is a crucial point to amplify the value of models beyond mere pictures \cite{bork19omilab}. One of the most prominent examples is the firing mechanism on Petri Nets \cite{reisig14petrinets}.
Also many domain-specific languages gain in value by the offered model operations. For example, 
model operations in the sense of model to model transformations play a crucial role in Model Driven Software Engineering \cite[chapter 8]{wimmer17mdse}.
Nevertheless, operations on models are often out of scope or simply ignored in formalizations. An exception is the theory of graph grammars and graph transformations \cite{taentzer10graphtransformation}. 
Formalisms based on logic are often critiqued for not being able to capture the operational syntax of modeling languages.
We argue that this is not an inevitable inability of these approaches and show some ideas how operations on models can also be supported by concepts from logic.

\subsubsection{Structural Events and Domain Events}
We adopt the notion of Olivé \cite[chapter 11]{olive07conceptual} who defines domain events, i.e. semantically and syntactically admissible operations on models, by decomposing them into the smallest possible changes in a model, the so called structural events. While Olivé only names deletion and insertion of objects and relations as structural events, for our purpose in the formalism at hand we also have to consider the change of attribute values as a third variant.

Given a language $\mathcal{L}$ we define $\mathcal{M}$ as the set of valid models, i.e. those language-structures fulfilling the constraints, and $\mathcal{M}^-$ as the set of all possible language constructs not necessarily complying to the postulates. Domain events are therefore mappings $de:\mathcal{M}\rightarrow\mathcal{M}$ whereas structural events are functions on arbitrary constructs: 

$$\text{Create}_\text{Type}: \mathcal{M}^- \rightarrow \mathcal{M}^-$$ 
$$\text{Set}^{obj}_\text{Attr}[value]: \mathcal{M}^- \rightarrow \mathcal{M}^-$$ 
$$\text{Delete}_\text{obj}: \mathcal{M}^- \rightarrow \mathcal{M}^-$$
Consider again model \texttt{M} of example \ref{petrinetmodelexample}. A valid domain event is the firing of the transition \textit{Serve}. This event is the concatenation of the three structural events of changing the attribute values 
$$Set^{Wait}_{tok}[1]\circ Set^{Idle}_{tok}[0]\circ Set^{Busy}_{tok}[1](\texttt{M})=\texttt{M}'.$$
None of these structural events alone is semantically valid, but together they form a semantically and syntactically admissible operation.
This also shows, that there are many structural events (we can set the attribute \textit{tokens} of each place to any number we like) but much less domain events.

Another point to be considered are pre- and postconditions of domain events. To capture these in a gene\-ric way we can use concepts from temporal logic \cite{kroeger2008}. With the logical operators from temporal logic we are able to formulate postulates considering both states of a model, before and after the application of a domain event, and define dependencies between both.

Concatenations of domain events form sequences of valid models $$\texttt{M}_0 \mapsto \texttt{M}_1 \mapsto \texttt{M}_2 \mapsto \cdots.$$ In Petri Nets for example the firing of a transition is a domain event. Therefore these sequences are of special interest as they reveal inaccessible states and final markings in a net when starting from a concrete model. This is closely related to the concept of marking graphs in Petri Nets \cite[Sec. 2.8]{reisig14petrinets}.

\subsection{Translators}
Another salient benefit of having an unambiguous and complete formalization of a modeling language is that it can serve as a single point of platform independent specification, thereby being precise enough to be automatically processed by a machine.
Of course a modeling language without a technical tool supporting the creation and execution of models is very much useless for the target audience.
When implementing a language many engineers have made the experience that available metamodeling platforms differ heavily in available concepts and functionality and thereby impose more or less severe restrictions on the final product \cite{kern11metameta}. 
So the implementation forces the engineer to think in the frame of the used platform and to modify the language to fit to the given meta$^2$model and available model processing algorithms. 
A further drawback of this current practice is the fact, that each effort of implementation is lost, whenever the language has to be transferred to another platform, may it be caused by missing functionality for new language features or a cessation of platform support.

With the formalization of a language as stipulated by the AMME lifecylce of modeling methods we derive a sort of platform independent code and close the gap between the specification document and the final implementation. By using the proposed formalism the specification of the main concepts is unified and therefore offers the possibility to be translated to any metamodeling platform.
Thus, the language specification stays on a platform-agnostic level and the complexity of the platform-specificity can be outsourced to a platform-specific translator. The feasibility of this endeavour has been shown by Visic et al. \cite{visic2016language,visic15mmdefinition}. When platforms change only the translator has to be adapted but not the platform-independent conceptualization of a language.

While Visic et al. stay at the level of translators of language syntax our attempt on the formalization of model operations shown in Section \ref{subsecop} holds promise to be able to integrate an automatic translation of the functionality of modeling languages.
The decomposition of domain events into the three types of structural events allows for an automatization of translating the modeling language specification to a concrete tool as most platforms offer methods for creating or deleting elements or changing attributes.

\section{Evaluation} \label{seceval}
To evaluate the proposed formalism we recap the requirements mentioned in the Section \ref{secrequirements}: 1) The formalism has to be complete regarding the general building blocks of a language, 2) it has to be faithful to the character of modeling languages as such, and 3) it must be generic in a way that it admits the formalization of any language.

The proposed formalism comprises the core concepts constituting a modeling language. These were chosen based on a survey by Kern et al. \cite{kern11metameta} and the concept discussion by Olivé \cite{olive07conceptual}. We restricted to the most common concepts, i.e. those appearing in at least half of the surveyed metamodeling platforms in \cite{kern11metameta}. In Section \ref{secformalmodelinglanguages} we also listed the concepts for future integration. Regarding the first requirement we conclude that the proposed definition of a modeling language is not yet complete, but depicts the most relevant core. This is also shown by the realizability of several case studies depicted in this paper.

In current scientific literature there is a consensus about modeling languages being formal languages by nature \cite{delcambre18referenceframework,guarino19,olive07conceptual,Partridge13,thalheim11chapter}. This supports our choice of using logic as basis for the formalism and underpins the adherence to the linguistic character of languages including the alphabet and the instantiation relation. 
Additional affirmation is given by the multitude of practical constructs and methods of formal language theory and its straightforward applicability to current research issues, which was exemplarily shown in Section \ref{secoutlookmethods}.

The generic realizability of arbitrary modeling languages is provided by construction, as the concepts integrated in the formalism stem from literature concerning conceptual modeling in general. A realization of the three divergent use cases in this paper and several more unpublished use cases conducted by the authors furthermore backs this claim.

The empirical evaluation of feasibility and usability so far has been main\-ly conducted via the realization of prototypical case studies of various domains (not all published). Three of them were shown in this paper.
Other cases guiding the advancement of the formalism are for example ER-diagrams starting in \cite{doeller18}. In the light of language interleaving we formalized a language for modeling smart cities \cite{bork15smartcities} besides the UML case study. To investigate the formalization of model operations we formalized Petri Nets and ProVis, a tool for math education providing sophisticated methods to process statistical diagrams \cite{doeller2020provis}.

A proof of concept for the significance of the presented formalism can be given by an implementation of translators to at least two different metamodeling platforms especially if we are able to integrate a specification of model operations. Such a tool is currently under design.

In parallel, a more outreaching empirical evaluation is currently being designed.
We will conduct a study with around sixty business informatics students in a university course about metamodeling. The students will be asked to apply the proposed formalism to the modeling language they develop during the course and to evaluate complexity and limitations. 

\section{Conclusion}
In this paper we presented a definition of modeling languages as formal languages $\mathcal{L}$ with a signature $\Sigma$ in the sense of logic.
The concept of a $\mathcal{L}$-structure canonically corresponds to the instantiation relation between model and language and led us to the definition of models as $\mathcal{L}$-structures.
To illustrate the specification of formal modeling languages we demonstrated the definition on the Petri Net modeling language. 
We applied the definition also on the meta level and developed \texttt{M2FOL} -- a formal modeling language for metamodels. 
\texttt{M2FOL} models are precise and complete and therefore we were able to show how to algorithmically derive a formal modeling language signature from its metamodel. \texttt{M2FOL} is self-describing, which can be seen by applying the algorithm to its own metamodel.

After the introduction of the formalism we gave an outlook to the potential and benefits of formalized modeling languages using the approach at hand. 
We addressed the topic of language interleaving and consistency. Established methods from formal language theory provide methods to create an interleaved formal language from existing ones. We illustrated the process on a case study using UML Class Diagrams and Sequence Diagrams.

Another topic with high potential for the automatization of language implementation is the formalization of model operations. We outlined how to break down algorithms on models in smallest possible building blocks able to be formalized. This allows model operations to become an integral part of the formal language specification. 

This formal specification -- syntax as well as operations -- precise enough to be processed by a machine yet platform-independent, additionally allows us to develop platform-specific translators, transferring the single source of language specification to realizations on different platforms.

With this common practice of defining metamodels and modeling languages, these languages become comparable, reusable, and open to modularization. 
To broad\-en the conceptual capabilities of our approach we will further investigate more subtle concepts to be integrated into the definition.
These are for example powertypes \cite[chapter 17.2]{olive07conceptual}, the concepts of mixins and extenders for modular metamodels as proposed in \cite{zivkovic2016}, or the structural types of relations identified in \cite{thalheim09}. 
For a practical application of the language \texttt{M2FOL}, a suitable tool for transforming graphical metamodels into formal ones will be developed.

Finally, by using a sophisticated mathematical theory as grounding for the definition of modeling languages we can use this knowledge stack as resource to further establish a formal foundation for modeling languages. We can investigate the subclass of conceptual modeling languages in the class of formal languages and approach old problems with new tools.

\bibliographystyle{splncs04}
\bibliography{../../library}

\end{document}